\documentclass[11pt]{article}

\usepackage{epsfig}
\begin{document}

\begin{center}
\bigskip

.\vskip1.5cm

{\LARGE Non-normal and Stochastic Amplification of Magnetic  \vskip0.1cm}

{\LARGE Energy in the Turbulent Dynamo: Subcritical Case}

\vskip1.0cm

{\large Sergei Fedotov}$^{1}$

\vskip 1.0cm

$^1$ Department of Mathematics, UMIST - University of Manchester

Institute of Science and Technology, Manchester, M60 1QD UK,

e-mail: Sergei.Fedotov@umist.ac.uk

Web-page: http://www.ma.umist.ac.uk/sf/index.html

\vskip0.5cm

Submitted to Phys. Rev. Lett.

\vskip1.0cm

\bigskip

\textbf{Abstract}\vskip0.3cm
\end{center}

Our attention focuses on the stochastic dynamo equation with non-normal
operator that gives an insight into the role of stochastics and
non-normality in galactic magnetic field generation. The main point of this
Letter is a discussion of the generation of a large-scale magnetic field
that cannot be explained by traditional linear eigenvalue analysis. We
present a simple stochastic model for the thin-disk axisymmetric $\alpha
\Omega $ dynamo involving three factors: (a) non-normality generated by
differential rotation, (b) nonlinearity reflecting how the magnetic field
affects the turbulent dynamo coefficients, and (c) stochastic perturbations.
We show that even for the \textit{subcritical case,} there are three
possible mechanisms for the generation of magnetic field. The first
mechanism is a deterministic one that describes an interplay between
transient growth and nonlinear saturation of the turbulent $\alpha -$effect
and diffusivity. It turns out that the trivial state is nonlinearly unstable
to small but finite initial perturbations. The second and third are
stochastic mechanisms that account for the interaction of non-normal effect
generated by differential rotation with random additive and multiplicative
fluctuations. In particular, we show that in the \textit{subcritical \ case }%
the average magnetic energy can grow exponentially with time due to the
multiplicative noise associated with the $\alpha -$effect.

\newpage

The generation and maintenance of large scale magnetic fields in stars and
galaxies has attracted enormous attention in past years \cite{Mof}-\cite
{RShS} (see also a recent review \cite{Widrow}). The main candidate to
explain the process of conversion of the kinetic energy of turbulent flow
into magnetic energy is the mean field dynamo theory \cite{KrR}. The
standard dynamo equation for the large scale magnetic field $\mathbf{B}(t,%
\mathbf{x})$ reads $\partial \mathbf{B/}\partial t=$ curl$(\alpha \mathbf{B}%
)+\beta \Delta \mathbf{B}+$curl$(\mathbf{u}\times \mathbf{B}),$ where $%
\mathbf{u}\ $is the mean velocity field, $\alpha \ $is the coefficient of
the $\alpha $-effect and $\beta $ is the turbulent magnetic diffusivity.
This equation has been widely used for analyzing the generation of the
large-scale magnetic field. Traditionally the mathematical procedure
consists of looking for exponentially growing solutions of the dynamo
equation with appropriate boundary conditions. While this approach has been
quite successful in the prediction of large scale magnetic field generation,
it fails to predict the \textit{subcritical} onset of a large-scale magnetic
field for some turbulent flow. Although the trivial solution $\mathbf{B}=0$
is linearly stable for the \textit{subcritical} case, the non-normality of
the linear operator in the dynamo equation for some turbulent flow
configurations leads to the transient growth of initial perturbations \cite
{nonnormal}. It turns out that the non-linear interactions and random
fluctuations might amplify this transient growth further. Thus, instead of
the generation of the large scale magnetic field being a consequence of the
linear instability of trivial state $\mathbf{B}=0$, it results from the
interaction of transient amplifications due to the non-normality with
nonlinearities and stochastic perturbations. The importance of the transient
growth of magnetic field for the induction equation has been discussed
recently in \cite{FI1, Proctor}. Comprehensive reviews of \textit{subcritical%
} transition in hydrodynamics due to the non-normality of the linearized
Navier-Stokes equation, and the resulting onset of shear flow turbulence,
can be found in \cite{Grossmann,SH}.

The main purpose of this Letter is to study the non-normal and stochastic
amplification of the magnetic field in galaxies. Our intention is to discuss
the generation of the large-scale magnetic field that cannot be explained by
traditional linear eigenvalue analysis. It is known that non-normal
dynamical systems have an extraordinary sensitivity to stochastic
perturbations that leads to great amplifications of the average energy of
the dynamical system \cite{F0}. Although the literature discussing the mean
field dynamo equation is massive, the effects of non-normality and random
fluctuations are relatively unexplored. Several attempts have been made to
understand the role of random fluctuations in magnetic field generation. The
motivation was the observation of rich variability of large scale magnetic
fields in stars and galaxies. Small scale fluctuations parameterized by
stochastic forcing were the subject of recent research by Farrell and
Ioannou \cite{FI1}. They examined the mechanism of stochastic field
generation due to the transient growths for the induction equation. They did
not use the standard closure involving $\alpha $ and $\ \beta $
parameterization. Hoyon with his colleagues has studied the effect of random
alpha-fluctuations on the solution of the kinematic mean-field dynamo\cite
{Hoyng}. However they did not discuss the non-normality of the dynamo
equation and the possibility of stochastic transient growth of magnetic
energy. Both attempts have involved only the linear stochastic theory.
Numerical simulations of magnetoconvection equations with noise and
non-normal transient growth have been performed in \cite{Proctor}

It is the purpose of this Letter to present a simple stochastic dynamo model
for the thin-disk axisymmetric $\alpha \Omega $ dynamo involving three
factors: non-normality, non-linearity and stochastic perturbations. Recently
it has been found \cite{Fedotov} that the interactions of these factors
leads to noise-induced phase transitions in a ``toy'' model mimicking a
laminar-to-turbulent transition. In this Letter we discuss three possible
mechanisms for the generation of a magnetic field that are not based on
standard linear eigenvalue analysis of the dynamo equation. The first
mechanism is a deterministic one that describes an interplay between linear
transient growth and nonlinear saturation of both turbulent parameters: $%
\alpha $ and $\beta $. The second and third are stochastic mechanisms that
account for the interaction of the non-normal effect generated by
differential rotation with random additive and multiplicative fluctuations.

Here we study the nonnormality and stochastic perturbation effects on the
growth of galactic magnetic field by using a Moss's ``no-z'' model for
galaxies \cite{Beck}. Despite its simplicity the ``no-z'' model proves to be
very robust and gives reasonable results compared with real observations. We
consider a thin turbulent disk of conducting fluid of uniform thickness $\
2h\ $and radius $\ R\ $($R\gg h$), which rotates with angular velocity $\
\Omega (r)$ \cite{ZRS,RShS}. We consider the case of $\ \alpha \Omega -$%
dynamo for which the differential rotation dominates over the $\alpha $%
-effect. Neglecting the radial derivatives one can write the stochastic
equations for the azimuthal, $B_{\varphi }\left( t\right) ,\ $and radial,$\
B_{r}\left( t\right) ,$ components of the axisymmetric magnetic field
\[
\frac{dB_{r}}{dt}=-\frac{\alpha (|\mathbf{B}|,\xi _{\alpha }(t))}{h}%
B_{\varphi }-\frac{\pi ^{2}\beta (|\mathbf{B}|)}{4h^{2}}B_{r}+\xi _{f}(t),
\]
\begin{equation}
\frac{dB_{\varphi }}{dt}=gB_{r}-\frac{\pi ^{2}\beta (|\mathbf{B}|)}{4h^{2}}%
B_{\varphi },\   \label{governing}
\end{equation}
where $\alpha (|\mathbf{B}|,\xi _{\alpha }(t))$ is the random non-linear
function describing the $\alpha -$effect, $\beta (|\mathbf{B}|)$ is the
turbulent magnetic diffusivity, $g=rd\Omega /dr\ $is the measure of
differential rotation (usually $rd\Omega /dr<0).$

Nonlinearity of the functions $\alpha (|\mathbf{B}|,\xi _{\alpha }(t))$ and $%
\beta (|\mathbf{B}|)$ reflects how the growing magnetic field $\mathbf{B}$
affects the turbulent dynamo coefficients. This nonlinear stage of dynamo
theory is a topic of great current interest, and, numerical simulations of
the non-linear magneto-hydrodynamic equations are necessary to understand
it. There is an uncertainty about how the dynamo coefficients are suppressed
by the mean field and current theories seem to disagree about the exact form
of this suppression \cite{backreaction}. Here we describe the dynamo
saturation by using the simplified forms \cite{Widrow}
\begin{equation}
\alpha (|\mathbf{B}|,\xi _{\alpha }(t)=(\alpha _{0}+\xi _{\alpha
}(t))\varphi _{\alpha }(|\mathbf{B}|),\;\;\;\beta (|\mathbf{B}|)=\beta
_{0}\varphi _{\beta }(|\mathbf{B}|),  \label{nonlinear}
\end{equation}
where $\varphi _{\alpha ,\beta }(|\mathbf{B}|)$ is a decaying function such
that $\varphi _{\alpha ,\beta }(0)=1.$ In what follows we use \cite{Widrow}
\begin{equation}
\varphi _{\alpha }(|\mathbf{B}|)=\left( 1+k_{\alpha }(B_{\varphi
}/B_{eq})^{2}\right) ^{-1},\;\;\;\varphi _{\beta }(|\mathbf{B}|)=\left( 1+%
\frac{k_{\beta }}{1+(B_{eq}/B_{\varphi })^{2}}\right) ^{-1},
\label{backreaction}
\end{equation}
\ where $k_{\alpha }$ and $k_{\beta }$ are constants of order one, and $%
B_{eq}$ is the equipartition strength. It should be noted that for the $%
\alpha \Omega -$dynamo the azimuthal component $B_{\varphi }\left( t\right) $
is much larger$\ $than the radial field$\ B_{r}\left( t\right) ,$ therefore,
$\mathbf{B}^{2}\simeq B_{\varphi }^{2}.$ We did not include the strong
dependence of $\alpha $ and $\beta $ on the magnetic Reynolds number $R_{m}$%
. The back reaction of the magnetic field on the differential rotation is
also ignored.

The multiplicative noise $\xi _{\alpha }(t)$ describes the effect of rapid
random fluctuations of $\alpha .$ We assume that they are more important
than the random fluctuations of the turbulent magnetic diffusivity $\beta $
\cite{Hoyng}. The additive noise $\xi _{f}(t)$ represents the stochastic
forcing of unresolved scales \cite{FI1}. Both noises are independent
Gaussian random processes with zero means $<\xi _{\alpha }(t)>=0,$ $<\xi
_{f}(t)>=0$ and correlations:
\begin{equation}
<\xi _{\alpha }(t)\xi _{\alpha }(s)>=2D_{\alpha }\delta (t-s),\;\;\;<\xi
_{f}(t)\xi _{f}(s)>=2D_{f}\delta (t-s).  \label{noise}
\end{equation}
The intensity of the noises is measured by the parameters $D_{\alpha }$ and $%
D_{f}$. One can show \cite{Fedotov} that the additive noise in the second
equation in (\ref{governing}) is less important.

The governing equations (\ref{governing}) can be nondimensionalized by using
an equipartition field strength $B_{eq},$ a length $h$, and a time $\Omega
_{0}^{-1},$ where $\Omega _{0}$ is the typical value of angular velocity. By
using \ the dimensionless parameters
\begin{equation}
g\rightarrow -\Omega _{0}|g|,\;\;\;\delta =\frac{R_{\alpha }}{R_{\omega }}%
,\;\;\;\varepsilon =\frac{\pi ^{2}}{4R_{\omega }},\;\;\;R_{\alpha }=\frac{%
\alpha _{0}h}{\beta },\;\;\;R_{\omega }=\frac{\Omega _{0}h^{2}}{\beta },
\end{equation}
we can write the stochastic dynamo equations in the form of SDE's
\[
dB_{r}=-(\delta \varphi _{\alpha }(B_{\varphi })B_{\varphi }+\varepsilon
\varphi _{\beta }(B_{\varphi })B_{r})dt-\sqrt{2\sigma _{1}}\varphi _{\alpha
}(B_{\varphi })B_{\varphi }dW_{1}+\sqrt{2\sigma _{2}}dW_{2},
\]
\begin{equation}
dB_{\varphi }=-(|g|B_{r}+\varepsilon \varphi _{\beta }(B_{\varphi
})B_{\varphi })dt\ ,  \label{basic}
\end{equation}
where $W_{1}$ and $W_{2}$ are independent standard Wiener processes. The
dynamical system (\ref{basic}) is subjected to the multiplicative and
additive noises with the corresponding intensities:
\begin{equation}
\sigma _{1}=\frac{D_{\alpha }}{h^{2}\Omega _{0}},\;\;\;\sigma _{2}=\frac{%
D_{f}}{B_{eq}^{2}\Omega _{0}}.  \label{intensity}
\end{equation}
It is well-known that the presence of noise can dramatically change the
properties of a dynamical system \cite{LH}. Since the differential rotation
dominates over the $\alpha $-effect ($R_{\alpha }\ll |R_{w}|),$ the system (%
\ref{basic}) involves two small parameters $\delta =$ $R_{\alpha }/R_{\omega
}$ and $\varepsilon =1/R_{\omega }$ whose typical values are $0.01-0.1$ ($%
R_{\omega }=10-100,$ $\ R_{\alpha }=0.1-1).$ These parameters play very
important roles in what follows. For small values $\delta $ and $\varepsilon
$ , the linear operator in (\ref{basic}) is a highly non-normal one $($ $%
|g|\sim 1).$ This can lead to a large transient growth of the azimuthal
component $B_{\varphi }\left( t\right) $ in a \textit{subcritical case.} We
then expect a high sensitivity to stochastic perturbations. Similar
deterministic low-dimensional models have been proposed to explain the
\textit{subcritical} transition in the Navier-Stokes equations (see, for
example, \cite{Trefethen,GS}). The main difference is that the nonlinear
terms in (\ref{governing}) are not energy conserving.

The probability density function $p(t,B_{r},B_{\varphi })$ obeys the
Fokker-Planck equation associated with (\ref{basic}) \cite{Gardiner}
\[
\frac{\partial p}{\partial t}=-\frac{\partial }{\partial B_{r}}\left[ \left(
\delta \varphi (B_{\varphi })B_{\varphi }+\varepsilon \varphi (B_{\varphi
})B_{r}\right) p\right] -\frac{\partial }{\partial B_{\varphi }}\left[
\left( |g|B_{r}+\varepsilon \varphi (B_{\varphi })B_{\varphi }\right) p%
\right] +
\]
\begin{equation}
(\sigma _{1}\varphi ^{2}(B_{\varphi })B_{\varphi }^{2}+\sigma _{2})\frac{%
\partial ^{2}p}{\partial B_{r}^{2}}.\
\end{equation}
Using this equation in the linear case one can find a closed system of
ordinary differential equations for the moments $<B_{r}^{2}>,$ $%
<B_{r}B_{\varphi }>,$ and $\ <B_{\varphi }^{2}>$
\begin{equation}
\frac{d}{dt}\left(
\begin{array}{c}
<B_{r}^{2}> \\
<B_{r}B_{\varphi }> \\
<B_{\varphi }^{2}>
\end{array}
\right) =\left(
\begin{array}{ccc}
-2\varepsilon  & -2\delta  & \sigma _{1} \\
-|g| & -2\varepsilon  & -\delta  \\
0 & -2|g| & -2\varepsilon
\end{array}
\right) \left(
\begin{array}{c}
<B_{r}^{2}> \\
<B_{r}B_{\varphi }> \\
<B_{\varphi }^{2}>
\end{array}
\right) +\left(
\begin{array}{c}
\sigma _{2} \\
0 \\
0
\end{array}
\right) .  \label{moments}
\end{equation}
The linear system of equations (\ref{moments}) allows us to determine the
initial evolution of the average magnetic energy $E(t)=<B_{r}^{2}>+<B_{r}B_{%
\varphi }>+<B_{\varphi }^{2}>.$ Similar equations emerge in a variety of
physical situations, such as models of stochastic parametric instability
that explain why the linear oscillator subjected to multiplicative noise can
be unstable \cite{BF}.

Now we are in a position to discuss three possible scenarios for the \textit{%
subcritical }generation of galactic magnetic field.

\textbf{Deterministic subcritical generation.} Let us examine the
deterministic transient growth of the magnetic field in the \textit{%
subcritical case.}To illustrate the non-normality effect consider first the
linear case without noise terms. The dynamical system (\ref{basic}) takes
the form
\begin{equation}
\frac{d}{dt}\left(
\begin{array}{c}
B_{r} \\
B_{\varphi }
\end{array}
\right) =\left(
\begin{array}{cc}
-\varepsilon  & -\delta  \\
-|g| & -\varepsilon
\end{array}
\right) \left(
\begin{array}{c}
B_{r} \\
B_{\varphi }
\end{array}
\right) .  \label{linear}
\end{equation}
Since $\delta <<1$, $\varepsilon <<1$ and $|g|\sim 1$, this system involves
a highly non-normal matrix. Even in the\textit{\ subcritical case (}$%
0<\delta <\varepsilon ^{2}/|g|$ see below\textit{) }when all eigenvalues are
negative, $B_{\varphi }$ exhibits a large degree of transient growth before
the exponential decay. Assuming that $B_{r}(t)=e^{\gamma t}$and $B_{\varphi
}(0)=b\ e^{\gamma t}$we find two eigenvalues $\gamma _{1.2}=-\varepsilon \pm
\sqrt{\delta |g|}$ (the corresponding eigenvectors are almost parallel). The
\textit{supercritical} excitation condition $\gamma _{1}>0$ can be written
as $\sqrt{\delta |g|}>\varepsilon $ or $\sqrt{R_{\alpha }R_{\omega }|g|}>\pi
^{2}/4$ \cite{RShS}. Consider the \textit{subcritical }case when $0<\delta
<\varepsilon ^{2}/|g|.$ The solution of the system (\ref{linear}) with the
initial conditions $B_{r}(0)=-2c\sqrt{\delta /|g|},$ $B_{\varphi }(0)=0\ $
is
\begin{equation}
B_{r}(t)=-c\sqrt{\frac{\delta }{|g|}}(e^{\gamma _{1}t}+e^{\gamma
_{2}t}),\;\;\;B_{\varphi }(t)=c(e^{\gamma _{1}t}-e^{\gamma _{2}t}).\
\end{equation}
Thus $B_{\varphi }(t)$ exhibits large transient growth over a timescale of
order $1/\varepsilon $ before decaying exponentially. In Fig. 1 we plot the
azimuthal component $B_{\varphi }$ as a function of time for $|g|=1,$ $%
\delta =10^{-4}$ and $\varepsilon =2\cdot 10^{-2}$ and different initial
values of $B_{r}$ ($B_{\varphi }(0)=0$).

\begin{figure}[tb]
\begin{center}
 \epsfxsize=0.6\textwidth \epsffile{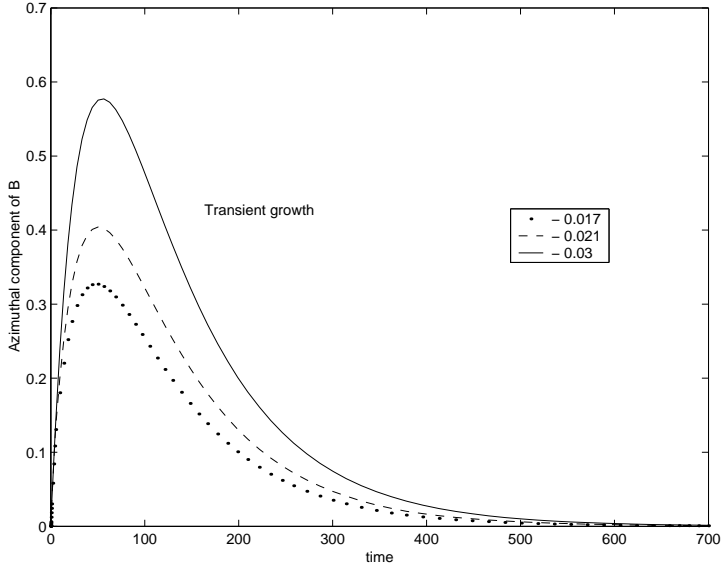}
\end{center}
 \caption{Linear case: the azimuthal component $B_{\protect\varphi }$ as a
function of time ($B_{\varphi }(0)=0$). for $|g|=1,$ $\delta %
=10^{-4}$ and $\varepsilon =2\cdot 10^{-2}$ and different initial
values of $B_{r}: - 0.017, - 0.021, - 0.03$.}
 \label{fig1}
\end{figure}

Of course without nonlinear terms any initial perturbation decays. However
if we take into account the back reaction suppressing the effective
dissipation ($\varphi _{\beta }(|\mathbf{B}|)$ is a decaying function), one
can expect an entirely different global behaviour. In the deterministic case
there can be three stationary solutions to (\ref{basic}). In Fig. 2 we
illustrate the role of transient growth and nonlinearity in the transition
to a non-trivial state using (\ref{backreaction}) with $k_{\alpha }=0.5$ and
$k_{\beta }=3$. We plot the azimuthal component $B_{\varphi }$ as a function
of time with the initial condition $B_{\varphi }(0)=0.$ We use the same
values of parameters $|g|,$ $\delta $ and $\varepsilon $ and three initial
values of $B_{r}(0)$ as in Fig. 1. One can see from Fig. 2 that the trivial
solution $B_{\varphi }=B_{r}=0$ is nonlinearly unstable to small but finite
initial perturbations of $B_{r},$ such as, $B_{r}(0)=-0.03$. For fixed
values of the parameters in nonlinear system (\ref{basic}), there exists a
threshold amplitude for the initial perturbation, above which $B_{\varphi
}(t)$ grows and below which it eventually decays.

\begin{figure}[tb]
\begin{center}
 \epsfxsize=0.6\textwidth \epsffile{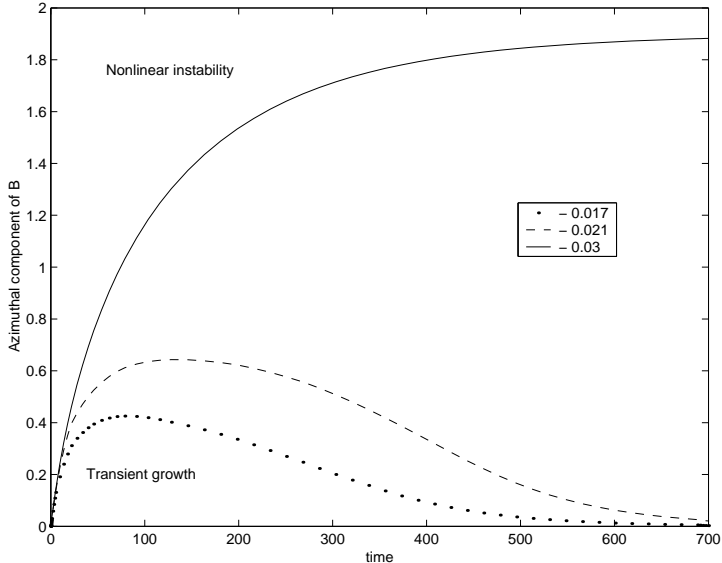}
\end{center}
 \caption{Nonlinear case: the azimuthal component $B_{\varphi }$ as a
function of time ($B_{\varphi }(0)=0$). for $|g|=1,$ $\delta %
=10^{-4}$ and $\varepsilon =2\cdot 10^{-2}$ and different initial
values of $B_{r}: - 0.017, - 0.021, - 0.03$.}
 \label{fig2}
\end{figure}

\textbf{Stochastic subcritical generation due to additive noise.} This
scenario has been already discussed in the literature \cite{FI1} (see also
\cite{F0} for hydrodynamics). The physical idea is that the average magnetic
energy is maintained by additive Gaussian random forcing representing
unresolved scales. It is clear that the non-zero additive noise ($\sigma
_{2}\neq 0$) ensures the stationary solution to (\ref{moments}). If we
assume for simplicity $\sigma _{1}=0$ and $\delta =0$ then the dominant
stationary moment is
\begin{equation}
<B_{\varphi }^{2}>_{st}=\frac{g^{2}\sigma _{2}}{4\varepsilon ^{3}}.
\label{stationary}
\end{equation}
We can see that due to the non-normality of the system (\ref{linear}) the
average stationary magnetic energy $E_{st}\sim <B_{\varphi }^{2}>_{st}$
exhibits a high degree of sensitivity with respect to the small parameter $%
\varepsilon :E_{st}\sim \varepsilon ^{-3}$ \cite{F0, Fedotov}.

\textbf{Stochastic subcritical generation due to multiplicative noise. }Here
we discuss the divergence of the average magnetic energy $%
E(t)=<B_{r}^{2}>+<B_{r}B_{\varphi }>+<B_{\varphi }^{2}>$ with time $t$ due
to the random fluctuations of the $\alpha -$parameter. Although the first
moments tend to zero in the \textit{subcritical case, }the average energy $%
E(t)$ grows as $e^{\lambda t}$ when the intensity of noise $\sigma _{1}$
exceeds a critical value. The growth rate $\lambda $ is the positive real
root of the characteristic equation for the system (\ref{moments})
\begin{equation}
(\lambda +2\varepsilon )^{3}-4\delta |g|(\lambda +2\varepsilon )-2\sigma
_{1}|g|=0.  \label{ch}
\end{equation}
For $\delta =0,$ the growth rate is $\lambda _{0}=-2\varepsilon +(2\sigma
_{1}|g|)^{1/3}$ as long as it is positive, and the excitation condition can
be written as $\sigma _{1}>\sigma _{cr}=4\varepsilon ^{3}/|g|.$ It means
that the generation of average magnetic energy occurs for $\alpha _{0}=0$ !
It is interesting to compare this criterion with the classical \textit{%
supercritical} excitation condition: $\delta |g|>\varepsilon ^{2}$\cite{RShS}%
. To assess the significance of this parametric instability it is useful to
estimate the magnitude of the critical noise intensity $\sigma _{cr}.$ First
let us estimate the parameter $\varepsilon =\pi ^{2}\beta /(4\Omega
_{0}h^{2}).$ The turbulent magnetic diffusivity is given by $\beta \simeq
lv/3,$ where $v$ is the typical velocity of turbulent eddy $v\simeq 10$ km s$%
^{-1},$ and $l$ is the turbulent scale, $l\simeq 100$ pc. For spiral
galaxies, the typical values of the thickness, $h,$ and the angular
velocity, $\Omega _{0},$ are $h\simeq 800$ pc and \ $\Omega _{0}\simeq
10^{-15}$ s$^{-1};$ $|g|\simeq 1$\cite{RShS}. It gives an estimate for $%
\varepsilon \simeq 3.2\times 10^{-2}$ , that is, $\sigma _{cr}\simeq
1.\,\allowbreak 3\times 10^{-4}.$ In general $\lambda (\delta )$ $=\lambda
_{0}+$ $(4/3|g|(2\sigma _{1}|g|)^{-1/3})\delta +$ $o(\delta ).$ This
analysis predicts an amplification of the average magnetic energy in a
system (\ref{basic}) where no such amplification is observed in the absence
of noise. The value of the critical noise intensity parameter $\sigma _{cr},$
above which the instability occurs, is proportional to $\varepsilon ^{3}$,
that is, very small indeed. To some extent, the amplification process
exhibits features similar to those observed in the linear oscillator
submitted to parametric noise \cite{BF}. To avoid the divergence of the
average magnetic energy, it is necessary to go beyond the kinematic regime
and consider the effect of nonlinear saturations.

In summary, we have discussed galactic magnetic field generation that cannot
be explained by traditional linear eigenvalue analysis of dynamo equation.
We have presented a simple stochastic model for the $\alpha \Omega $ dynamo
involving three factors: (a) non-normality due to differential rotation, (b)
nonlinearity of the turbulent dynamo $\alpha $ effect and diffusivity $\beta
$, and (c) additive and multiplicative noises. We have shown that even for
the \textit{subcritical case,} there are three possible scenarios for the
generation of large scale magnetic field. The first mechanism is a
deterministic one that describes an interplay between transient growth and
nonlinear saturation of the turbulent $\alpha -$effect and diffusivity. We
have shown that the trivial state $\mathbf{B}=0$ can be nonlinearly unstable
with respect to small but finite initial perturbations. The second and third
are stochastic mechanisms that account for the interaction of non-normal
effect generated by differential rotation with random additive and
multiplicative fluctuations. We have shown that multiplicative noise
associated with the $\alpha -$effect leads to exponential growth of the
average magnetic energy even in the \textit{subcritical case. }

\textbf{Acknowledgements} I am grateful to Anvar Shukurov for constructive
discussions.


\begin{thebibliography}{99}
\bibitem{nonnormal}  An operator is said to be non-normal if it does not
commute with its adjoint in the corresponding scalar product. Typical
examples of transient growth can be represented by the functions: $%
f_{1}(t)=t\exp (-t)$ and $f_{2}(t)=$ $\exp (-t)$ $-\exp (-2t)$ for $t\geq 0.$
\end{thebibliography}
\end{document}